# THz FIELD CONTROL OF IN-PLANE ORBITAL ORDER IN $La_{0.5}Sr_{1.5}MnO_4$

Timothy A. Miller[1], Ravindra W. Chhajlany[1,2], Luca Tagliacozzo[1], Bertram Green[3], Sergey Kovalev[3], Dharmalingam Prabhakaran[4], Maciej Lewenstein[1,5], Michael Gensch[3]*, Simon Wall[1]*

1 ICFO-Institut de Ciències Fotòniques, Av. Carl Friedrich Gauss 3, 08860 Castelldefels, Barcelona, Spain
2 Faculty of Physics, Adam Mickiewicz University, Umultowska 85, 61-614 Poznań, Poland
3 Helmholtz-Zentrum Dresden Rossendorf, Bautzner Landstraße 400, 01328 Dresden, Germany
4 Department of Physics, Clarendon Laboratory, University of Oxford, Oxford, OX1 3PU, United Kingdom
5 ICREA-Institució Catalana de Recerca i Estudis Avançats, Lluís Company 23, 08010 Barcelona, Spain

## ABSTRACT

*In-plane anisotropic ground states are ubiquitous in correlated solids such as pnictides, cuprates and manganites. They can arise from doping Mott insulators and compete with phases such as superconductivity; however their origins are debated. Strong coupling between lattice, charge, orbital and spin degrees of freedom results in simultaneous ordering of multiple parameters, masking the mechanism that drives the transition. We demonstrate that the orbital domains in a manganite can be oriented by the polarization of a pulsed THz light field. Through the application of a Hubbard model, we show that domain control can be achieved by enhancing the local Coulomb interactions which drive domain reorientation. Our results highlight the key role played by the Coulomb interaction in the control and manipulation of orbital order in the manganites and demonstrate a new way to use THz to understand and manipulate anisotropic phases in a potentially broad range of correlated materials.*

## INTRODUCTION

The simultaneous ordering in multiple degrees of freedom is a general phenomenon in layered correlated solids. These ordered states are anisotropic and break the four-fold in-plane symmetry found in the high-temperature phase. In pnictides spin, charge, orbital, and structural degrees of freedom lock together at the nematic transition[1]; in the manganites charge, orbital, and structural degrees of freedom simultaneously order to give rise to orbital phases; similar transitions occur in cuprates[2] and nickelates[3]. Understanding which degree of freedom is primarily responsible for these transitions is a focus of condensed matter physics[1] and may help to explain why such phases are favoured over alternatives such as superconductivity. Furthermore, dynamic control of the anisotropy, which strongly influences the material properties[4], could lead to novel devices based on correlated materials.

$La_{0.5}Sr_{1.5}MnO_4$ (LSMO) is a prototypical manganite. The high-temperature state is a paramagnetic semiconductor. The crystal has a layered two dimensional structure of manganese planes as shown in Figure 1a. Below the orbital ordering temperature, $T_{OO}$ = 230 K, electrons in the 2-fold degenerate $e_g$ level of the Mn ions localize on alternating Mn sites. Simultaneously, the occupied $e_g$ orbitals align. This second-order phase transition breaks the in-plane symmetry, and two possible isoenergetic orbital domains can form. The domains consist of CE-type[5] zig-zag chains[6] oriented along one of two possible crystallographic directions which give rise to anisotropic optical and electronic properties[7,8]. Below $T_N$ = 110 K, antiferromagnetic order also emerges.

The origin of orbital order remains debated, and electronic correlations[9–11], structural distortions[12–15], or spin interactions[16–18] have each been suggested as the primary factor. Our novel experimental technique and theoretical modelling address this problem by focusing on the switching of one domain type to the other under an applied THz field observed using optical birefringence. We show that the multi-site Coulomb interaction provides the energy for domain reorientation, even when considering structural distortions. For the single-layer manganite LSMO studied here, field-assisted hopping of electrons from two highly occupied-sites onto the same less-occupied site increases Coulomb repulsion and causes domain reorientation when the THz field is orthogonal to the CE chain direction. Our model also suggests that structural distortions create preferential electron hopping directions, which modify Coulomb interactions and may cause domain reorientation in other systems. As Coulomb interactions ultimately drive domain switching regardless of the field coupling method our results suggest that electronic corrections are a key factor in the formation of orbital order.

## Results

### Optically Probing Orbital Order

We harness the anisotropic reflectance to measure orbital alignment. The polarization of an incident helium-neon laser (HeNe) beam is oriented to lie between the two zig-zag chain directions, A and B (Figure 1b). Reflected light incident on a domain of Type A experiences a polarization rotation of opposite direction to that experienced by light incident on a domain of Type B. As the HeNe beam is much larger than the orbital domain size[19], the spatially averaged reflected beam does not experience a net rotation when equal amounts of both domains are present (Figure 1b). However, both domains have a non-zero projection onto the $R_\perp$ axis; thus the intensity of $R_\perp$ as a function of temperature allows us to observe the onset of orbital order, while $R_A - R_B$ measures the preferential formation of one domain type (Figure 1c).

### THz Field Induced Birefringence

Figure 2a shows the experimental setup for detecting THz induced domain alignment. 10-ps-long THz pulses are generated at 13 MHz by a novel infra-red free-electron laser (FEL)[20]. The polarization of the THz field is controlled by a mechanically rotating half-wave plate so that the polarization periodically lies along either domain direction A or B. Variations in the polarization rotation of a HeNe probe beam are detected at the rotation rate of the THz polarization using a lock-in amplifier. This enables the measurement of small changes in the probe polarization whilst maintaining a constant isotropic heat load on the sample (see methods for details). We verified that the measured signal did not change when the rotation rate of the waveplate was changed from 1 – 4 Hz.

Figure 2b shows the THz-induced anisotropy as a function of THz fluence for a sample temperature of 170 K, below $T_{OO}$ but above $T_N$. The THz-induced anisotropy signal was observed to grow linearly with the field intensity for several different excitation wavelengths, and saturation occurred at the highest intensities. By using the literature values for the anisotropy of a single domain[8], we calculate that we could align approximately 10% of the domains before saturation and heating effects dominate. Figure 2c shows that in this resonance-free region the induced anisotropy was weakly dependent on the THz wavelength. Due to the finite bandwidth of the FEL pulses, longer wavelengths have a longer pulse duration and lower peak fields. Thus, as the ability to align domains decreases slightly for longer wavelengths, we conclude that the peak field is more important for domain alignment than the total energy in the THz pulse, which remains constant.

To verify that the observed anisotropy signal was due to domain alignment by the field polarization, we performed several checks. Figure 2d shows the anisotropic signal as a function of the input HeNe probe polarization angle. As expected, when the probe polarization is rotated by 45 degrees to align with an orbital domain axis the polarization state is not modulated by domain alignment, and a further rotation by 45 degrees reverses the polarity of the signal. Figure 2e shows that a largely reduced signal was observed when an additional quarter-wave plate was inserted into the THz beam before the rotating half-wave plate to produce rotating circularly polarized light. The remaining signal is due to a small residual ellipticity in the polarization. This demonstrates the significance of the THz polarization direction and allows us to exclude pointing variations in the THz beam or thermal effects as the primary source of the anisotropy signal.

To understand these observations, we measure a detailed power and temperature dependence. Figure 3a shows the temperature and power dependence of power-normalized THz-induced net anisotropy. On cooling below $T_{OO}$, the anisotropic signal increases rapidly for moderate THz powers. This occurs in step with the total orbital order shown in Figure 1c. This rapid increase is due to the increasing correlation length of domains that occurs as the sample cools. On further cooling, the induced anisotropy saturates at approximately 200 K and then starts to diminish, with the signal almost gone below $T_N$.

For higher THz powers the onset for anisotropy signal is shifted to lower temperatures. This shift can be identified as the result of DC heating, with 2 W of THz power inducing a heating of 85 K. The thermal origin of this shift is clearly seen in Figure 3b, which shows the temperature dependence of induced anisotropy for different THz powers. At the lowest powers, identical power-normalized temperature dependencies are observed, demonstrating the linear and non-thermal nature of the signal. The high power signal shows a large thermal shift. If this shift is compensated, the same power-normalized signal is observed. The shift in the rising edge can be used to determine the power dependence of the thermal contribution, which is found to scale as $I_{THz}^3$ (white dotted line) and thus only contributes at the highest powers.

## Measured THz Field Coupling Mechanism

The above results clearly demonstrate that the THz field can be used to align orbital domains, but the question remains as to how this is achieved. Thermodynamically, the energy of a dielectric material in an external field is proportional to $\varepsilon(\omega)E^2$ where $\varepsilon(\omega)$ is the dielectric function at the frequency of the THz field. An anisotropy in the dielectric function at the THz driving frequency would create an energy difference between domains aligned parallel or perpendicular to the THz field. This energy difference is given as $\Delta\varepsilon(\omega)E^2$, where $\Delta\varepsilon(\omega)$ is the difference between the dielectric function along the different domain directions. Anisotropic dielectric functions can arise as a direct result of the order parameter, as in ferroelectrics, where the polarization modifies the linear dielectric function, or indirectly, as in ferromagnets with magnetostriction, where the magnetic-field-induced strain induces the optical anisotropy.

Although both cases have the potential for field-induced alignment, the switching mechanism will be very different. In the latter case the electric field does not couple to the order parameter directly and instead shifts the thermodynamic potential to favour one domain orientation. Thermal fluctuations could move the material towards the more favourable orientation. Such a mechanism is unlikely to occur in our experiment as we use pulsed fields which are only 'on' for approximately 10 ps. This makes this indirect process extremely unlikely as the majority of the time the sample is not in an applied field and the two domains are energetically degenerate.

Direct manipulation of the order parameter by THz fields has been achieved through a linear coupling of the field to resonant electric[21] or magnetic[22] dipoles. However, to date, domain control with THz fields has not been achieved experimentally. Furthermore orbital ordering is not described by a macroscopic polarization or magnetization; thus a linear coupling of the THz field to the order parameter cannot provide the driving force for domain re-orientation in LSMO.

## HUBBARD MODEL HAMILTONIAN

To investigate how the THz field can still couple to orbital order, we consider a phenomenological 2D extended Hubbard model for the $e_g$ orbitals $|\infty\rangle = d_{3x^2-r^2}$ and $|8\rangle = d_{3y^2-r^2}$. The Hamiltonian, $H$, can be expressed as

$$H = -t\left(\sum_i c^\dagger_{i,\infty} c_{i+\mathbf{e}_x,\infty} + c^\dagger_{i,8} c_{i+\mathbf{e}_y,8} + h.c.\right) + U \sum_i n_{i,\infty} n_{i,8} + V \sum_{<i,j>} n_i n_j \qquad (1)$$

where U and V are the on-site and nearest neighbour Coulomb interaction terms respectively, $n_{i,\alpha}$ denotes the occupation of the orbital $\alpha = \infty, 8$ and $n_i = n_{i,\infty} + n_{i,8}$ is the total occupancy of site $i$. The spatial anisotropy of the d-orbitals plays a key role in the manganites[6], and for simplicity we take the limit that electrons can only hop between orbitals of the same character. The hopping between manganese ions decreases the energy of the electrons by $t$ and occurs via a bridging oxygen 2p orbital. The hopping probability depends on $\theta$, the Mn-O-Mn bond angle, as $t = t_0 \cos(\pi - \theta)^2$ where $t_0$ is the value of the hopping for an undistorted straight bond[23].

Figure 4a shows the computed energy diagram for different orbital configurations as a function of the onsite Coulomb energy $U$. As we are studying the state of the system above the magnetic ordering temperature, the spin state is ignored. In this case, with $t \ll U, V$, the experimentally-observed CE-type orbital ordering ground state is obtained from a 4$^{th}$ order perturbative expansion of the Hamiltonian, the details of which can be found in Supplemental Note 1 and are summarized in Supplemental Table 1. As expected, the two types of CE-domains are energetically degenerate.

Previous studies on control of magnetic and orbital ordering in LSMO have focused on order melting from optical[24–26] or vibrationally resonant mid-IR pulses[27,28]. The former predominantly influences the charge system whereas the latter modulates the lattice. Optical excitation triggers charge transfer between Mn sites, directly perturbing the orbital order. Resonant mid-IR light couples to IR active phonon displacement that directly modulates the Mn-O-Mn bond angle[29], which modulates the electronic hopping energy, $t$. We now consider how these two mechanisms can break the degeneracy between the two lowest energy domains obtained from Eqn. 1.

## COMPETING COUPLING MECHANISMS

First we consider the influence of the field directly on the charge degree of freedom. The THz photon energy is insufficient to trigger charge transfer excitations but instead polarizes the electronic orbitals along the field polarization direction. The effect of the THz field can then be treated quasi-statically in the length gauge as

$$H_E = -\mathbf{E} \cdot \sum_i \sqrt{2}\, r n_i, \qquad (2)$$

where $r$ is the Mn-Mn bond length. We calculate the energy difference of the ground state when the field is applied along or perpendicular to the chain direction and find that the degeneracy of the orbital domains is split. The energy splitting was found to scale as

$$\Delta\rho_\text{F} = \rho_{//} - \rho_\perp = -2t^4 \frac{(476U^3+5337U^2V+15348UV^2-7036V^3)}{30375V^5(U+4V)^3}(\sqrt{2}r)^2 E^2 \tag{3}$$

Domains where the field is applied along the chain direction become more stable than those with the field perpendicular to the chains. The model recovers an energy splitting that scales as $E^2$, matching the experimental observation. Details of the calculation are given in Supplementary Note 2.

Secondly we consider how the structure distortions induced by the ions in response to the electric field could induce domain alignment. As the oxygen and manganese ions are oppositely charged, they will move in opposing directions in response to the applied field, modifying the Mn-O-Mn bond angle and thus the hopping probability, $t$.

There is some controversy over the crystal structure of LSMO, with some groups reporting the I4/mmm structure above and below $T_\text{OO}$[30] and others reporting of a structural transition to a Cmmm phase below $T_\text{OO}$[31]. In the high symmetry I4/mmm phase the Mn-O-Mn bond angle is $\theta = 180°$. As shown in Figure 4b, the result of ionic motion in this phase uniformly decreases all bond angles in the same way. This renormalizes $t$ to a lower value, but this does not split the degeneracy of the domain energies. Thus we do not expect to be able to control orbital domains in LSMO through structural distortions induced by the electric field in the I4/mmm phase or in any other material with linear bonds at equilibrium.

However in the Cmmm space group, the Mn-O-Mn bond angle is decreased to 176.72° at equilibrium. This distortion is, in general, larger in the cubic manganites. In this case, Figure 4b shows that two bond angles are increased towards the ideal 180° angle and two bond angles are decreased under an applied field. Assuming small changes in bond angle, the change in hopping term becomes $t_\pm = t(1 \pm aE \tan \theta)$, where $aE$ is the field-induced change in the bond angle $\theta$, which can be obtained from considering the response of a dipole in an electric field (see Supplementary Note 3). Eqn. 1 can be re-evaluated with the alternating hopping factors. The resulting domain splitting is given by

$$\Delta\rho_\text{t} = \rho_{//} - \rho_\perp = 128t^4(a \tan \theta)^2 E^2 \frac{7U-2V}{135V^3(U+4V)} \tag{4}$$

The domain with lower energy is again the domain parallel to direction of the applied field, and the domain energy difference still scales with the intensity of the electric field.

## DISCUSSION

The domain switching processes that arise from these models are depicted in Figure 5. In the field driven case (a), the electric field pushes the charges between the Mn sites. When the field is aligned perpendicular to the domain chain direction, charges from two $Mn^{3+}$ sites are forced on the same $Mn^{4+}$ site. This increases the onsite Coulomb energy penalty and acts as the force to re-orientate the domain. If the orbital flips, the new domain structure is generated. In this case every $Mn^{3+}$ charge is moved onto a separate $Mn^{4+}$ site and thus does not experience the extra Coulomb repulsion.

A very similar process occurs when the domains are aligned due to changes in the tolerance factor (Figure 5b). Here, when the field is aligned perpendicular to the domain direction, the hopping probability is improved such that charge from two $Mn^{3+}$ sites is again moved onto the same $Mn^{4+}$. Like before, the Coulomb energy penalty acts as the force to reorient the domain by minimizing the energy penalty.

By considering the relative strength of the two processes we can determine the most likely driving mechanism in LSMO. Using values of *U* = 5 eV, *V* = 1 eV and *t* = 0.25 eV, based on ref [32], and the appropriate bond angles for LSMO in the Cmmm phase, we find that $\left|\Delta\rho_t / \Delta\rho_F\right| \approx 1 \times 10^{-3}$. Thus this simple model predicts that field induced alignment is more likely for LSMO. Structural control is most effective when IR displacements can induce large anisotropy in the hopping term. This occurs when the system starts in a state with large equilibrium bond angles, enabling larger angular changes for the same oxygen displacement. These displacements may be further enhanced by using fields resonant with the Mn-O-Mn bond. Thus the structural distortion mechanism may dominate in the cubic manganites, such as $Pr_{0.7}Ca_{0.3}MnO_3$, where there are significantly larger initial Mn-O-Mn bond angles.

The model can also qualitatively explain the experimentally observed decrease in domain alignment efficiency when the system magnetically orders. Below $T_N$ the spins order ferromagnetically along the chains, but antiferromagnetically across the chains[6]. Although perpendicularly-oriented domains are still energetically less favourable than parallel ones in the presence of the field, electrons that are transferred across the chains experience an additional energy penalty due to Hund's rule, and domains can only switch if the localized $t_{2g}$ spins also rotates. As the field does not couple to the spin degree of freedom directly, this process only happens due to random thermal fluctuations which are unlikely to occur during the short period of time in which the electric field is applied. Nearest-neighbour spin correlations appear well above $T_N$ and thus can start to impede domain alignment before long range spin order sets in[33]

Our results establish that pulsed THz fields can be used to *control* the anisotropy axis of manganites and represent the first demonstration of THz non-contact control over domains, moving beyond previous studies which examined melting of orbital order by THz fields. Our model shows the importance of Coulomb interactions as a driving mechanism for domain rotation in the manganites and strongly suggests that domain alignment is driven by field-induced hopping in manganites with low structural distortions. Combining the techniques presented here with X-ray based imaging experiments[34] would further verify the domain switching process and enable observation of the domain switching pathway and speed.

Similar experiments should also be applied to other anisotropic phases in the cuprates and pnictides to determine the relative strength of the Coulomb and structural mechanisms in these materials. Furthermore, control of orbital occupation is an emerging method for controlling electronic properties which has typically been achieved with strain or charge transfers from interface states[35]. Our results show that orbital control can be achieved dynamically with THz fields, opening new ways to control device properties and suggest orbital domains could be an alternative to magnetic data storage.

## METHODS

### BIREFRINGENCE MEASUREMENT

Figure 1c shows the experimental setup for detecting THz-induced domain alignment. A single crystal of LSMO was grown using the float-zone method, C-cut, optically polished, and mounted in a cryostat with quartz windows. Intense THz light and 632nm light from a continuous-wave HeNe laser were incident on the LSMO sample at near-normal incidence (<0.5deg). The polarization characteristics of the HeNe beam were analysed as a function of THz polarization.

Intense, tuneable, linearly-polarized pulses of THz-radiation with Fourier limited pulse durations between 10 and 15 ps are generated at a 13 MHz repetition rate from the FELBE Free Electron Laser[20]. The THz light was focused to a diffraction limited spot onto the sample using an off-axis parabolic mirror with through-hole to permit HeNe co-linearity. The THz polarization was rotated by a mechanical half-wave plate mounted in a motorized rotation stage set to rotate at 4 Hz. This causes a THz polarization rotation at 8Hz. To check the effect of circularly polarized THz, a fixed quarter-wave plate was placed upstream of the rotating half-wave plate to give rotating circularly polarized THz. This modulation technique keeps the thermal load on the sample constant, while at the same time enabling the detection of the small domain alignment signal. Powers were measured with a calibrated power meter and the fluences reported were calculated by taking the pulse energy, obtained by dividing the power by the repetition rate of the FEL, and dividing by the spot size. THz light was attenuated upstream of the experiment using calibrated attenuators provided by the FEL facility. The THz spot sizes were measured introducing a flip-mirror between the sample and the focusing off axis parabolic mirror and imaging on a pyroelectric camera (SPIRICON Pyrocam IIIHR).

Linearly polarized HeNe light was used to probe the effect of the THz on the orbital order. The polarization was chosen to lie between the two orbital directions such that there was a non-zero polarization projection into both orbital domains. The HeNe beam was focused to a spot much smaller than that of the THz on the sample, and the reflection was collected. The reflected beam passed through a beamsplitter to provide a reference of the total reflected intensity. The remaining HeNe light passed through a Wollaston prism to separate the light into Type A and Type B components and made incident on a balanced detector. The signal from the balanced detector was fed into a lock-in amplifier triggered at twice the frequency of the THz polarization rotation. This is appropriate as positive and negative THz electric fields produce the same force for domain rotation. The lock-in signal was measured over 10 s of acquisition time, and the mean value was used as the induced anisotropy. To normalize the wavelength-dependence of the THz spot size, the induced-anisotropy signal was normalized by the wavelength of the THz squared.

## CONFLICT OF INTEREST STATEMENT

The authors declare that there are no conflicting financial interests with this work.

## CORRESPONDING AUTHORS

Email: simon.wall@icfo.es, m.gensch@hzdr.de

## AUTHOR CONTRIBUTIONS

SW conceived of the experiment. TAM, BG, SK, MG, and SW planned and designed the experimental setup and performed measurements. SW analysed the data. DP provided characterized samples. Theoretical work was performed by RWC, LT and ML. All authors participated in writing the manuscript.

## ACKNOWLEDGEMENTS

SW acknowledges fruitful discussions with A. Cavalleri and P. Kirchmann and financial support from Ramon y Cajal program RYC-2013-14838 and Marie Curie Career Integration Grant PCIG12-GA-2013-618487. TAM is supported by the COFUND action of the European Commission and the Severo Ochoa Program of the Spanish Ministry of Economy and Competitiveness. R.W.C. acknowledges a Mobility Plus fellowship from the Polish Ministry of Science and Higher Education and the (Polish) National Science Center Grant No. DEC-2011/03/B/ST2/01903. R.W.C, L.T and M.L acknowledge support of Foundation of Polish Science, Spanish Government Grant FOQUS (FIS2013-46768), ERC AdG OSYRIS, EU IP SIQS, EU STREP EQuaM and EU FET Proactive QUIC. Part of this work has been supported by the European Commission through the CALIPSO project under the EC contract 312284.

## FIGURES

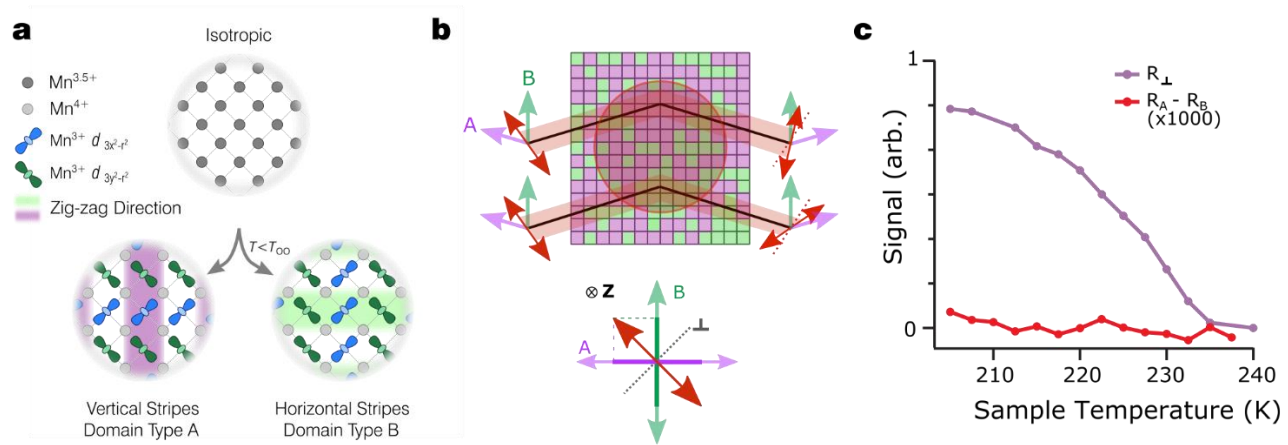

**Figure 1 | Orbital domain structure in $La_{0.5}Sr_{1.5}MnO_4$. a** Orbital ordering in LSMO. Above $T_{oo}$ = 230 K, the 2D Mn planes are isotropic with a nominal Mn valence of +3.5. On cooling LSMO adopts a CE-type charge and orbital ordering with the two orbital domains shown. The zig-zag stripes show the internal domain anisotropy direction, and the domains are related by a 90° rotation. The orbitals shown are the two $e_g$ orbitals of the Mn ion. **b** The optical method used to probe orbital order. (top) Light (red arrow) polarized between the domain directions (purple/green arrows) is rotated left or right upon reflection depending on the domain probed. Samples adopt a multi-domain structure on cooling and the HeNe probe measures multiple domains. (bottom) Any rotation of the incident light by either domain type will produce a projection in the perpendicular polarization, $R_\perp$. Measurement of $R_\perp$ therefore measures the presence of orbital domains without distinguishing between domain types. **c** Measure of orbital order and domain structure in LSMO. By measuring the perpendicular component of the reflected light, $R_\perp$, the orbital order can be measured. Measurements of $R_A$-$R_B$ give the net domain alignment, or net optical anisotropy. No $R_A$-$R_B$ signal is observed without incident THz light, indicating the sample is in a multi-domain state with equal population of both domain types.

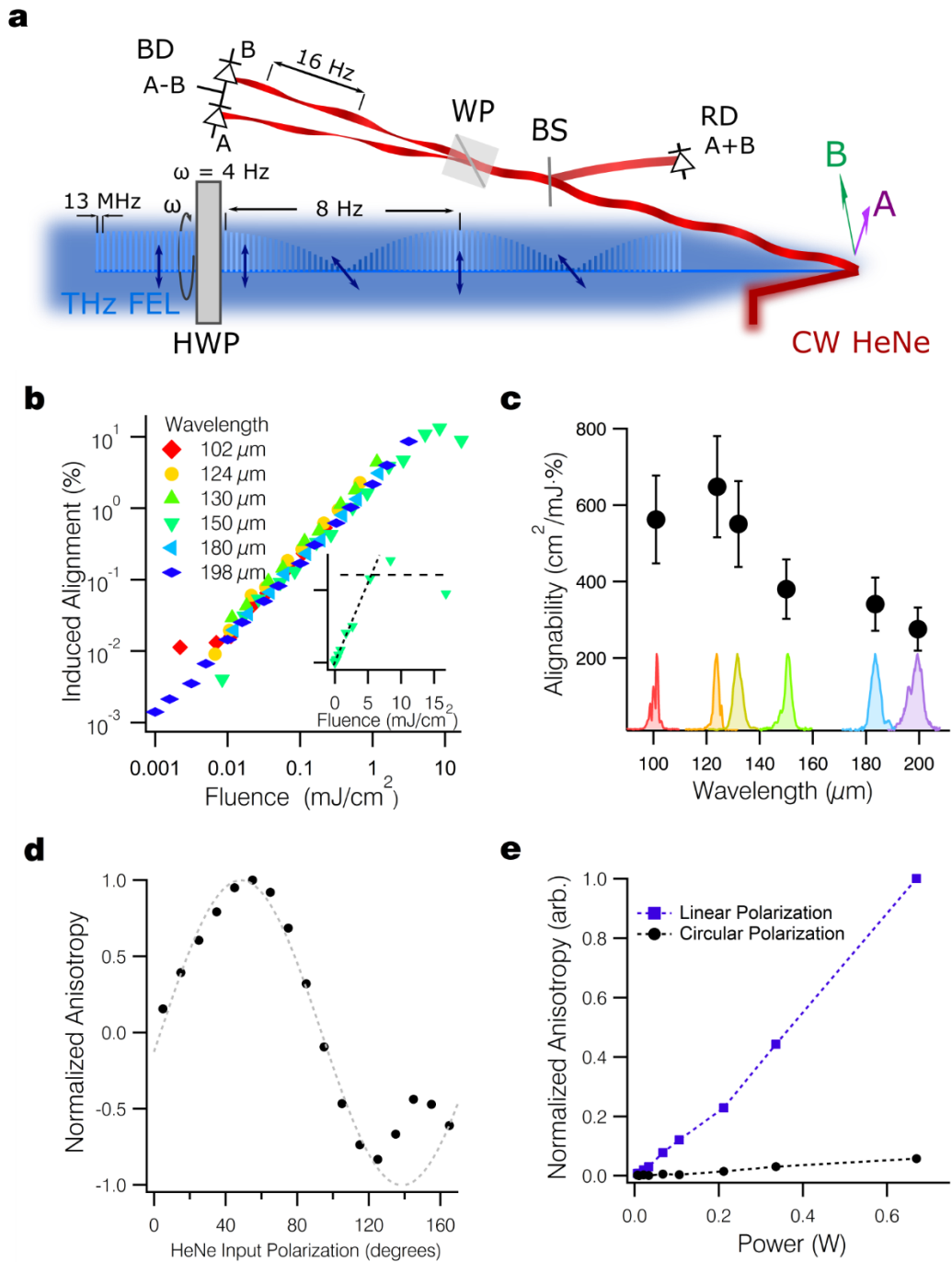

**Figure 2 | THz-induced alignment of orbital domains. a** 10 ps pulses of linearly polarized THz radiation are generated at 13 MHz from a free-electron laser. The radiation passes through a half-wave plate (HWP) rotating at a frequency $\omega$ = 4 Hz in the plane normal to the beam propagation. This rotates the THz polarization at 2 $\omega$. The THz is focused onto the sample, which is mounted in a cryostat, through a c-cut quartz window at normal incidence. A continuous wave (CW) HeNe laser arrives collinearly with the THz. Some of the reflected beam is separated by a beam splitter (BS) and recorded by a reference diode (RD) to normalize for changes in total reflectivity from the sample. The remaining reflected HeNe light is split by a Wollaston prism (WP) and collected on a balanced detector (BD) to give a signal which is sensitive to the net optical anisotropy ($R_A$-$R_B$). Changes in the anisotropy are detected with a lock-in amplifier at twice the frequency of the THz polarization rotation. **b** A

log-log plot shows that the power dependence of the anisotropy change is roughly linear before saturating at approximately 10%. The insert shows the same data on a linear plot for 150 µm THz light. The data is for a sample temperature of 170 K. **c** Induced anisotropy per unit fluence for each of the probe wavelengths. The spectra of the incident THz light are also shown. Error bars show one standard deviation. **d** The dependence of the measured THz induced anisotropy as a function of probe polarization showing the expected angular dependence. **e** THz polarization dependence of the induced anisotropy. The change in anisotropy with circularly polarized THz is substantially reduced and the remaining signal results from a small residual linearly polarized component of the THz field.

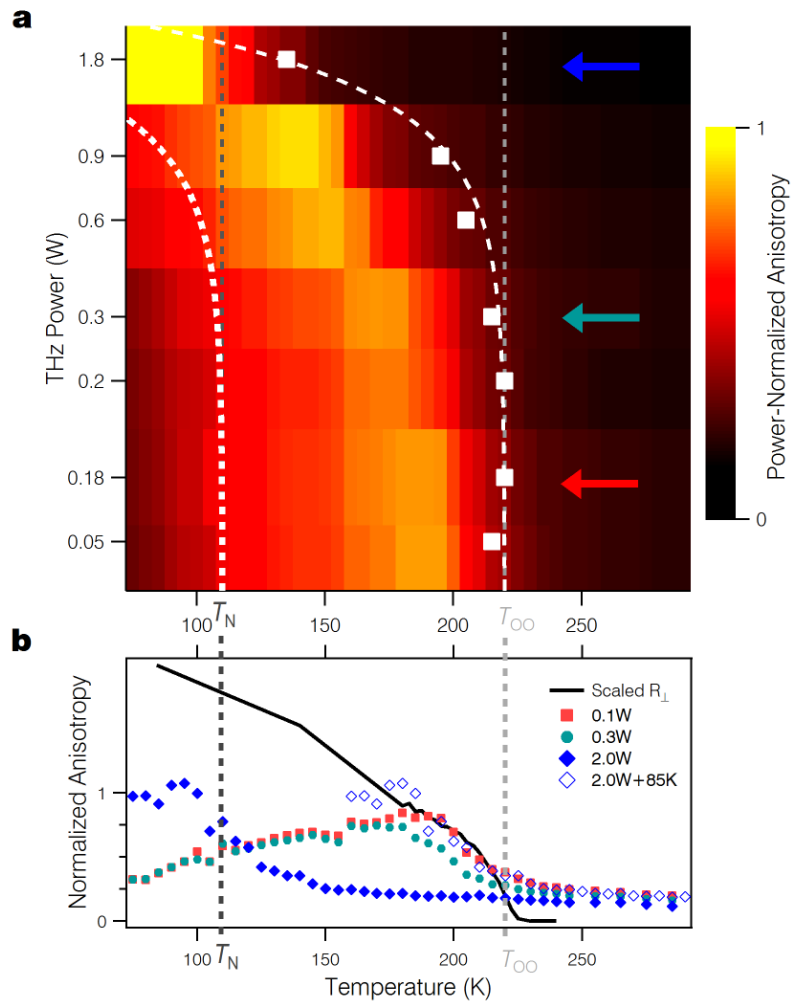

**Figure 3 | Temperature dependence of THz-induced orbital domain alignment.** The 2D plot, **a**, shows the THz induced anisotropy signal when normalized by the incident power. Dark dashed lines correspond to the orbital, $T_{OO}$, and magnetic, $T_N$, ordering temperatures. White squares correspond to the onset of the THz induced signal. THz alignment is only observed for $T<T_{OO}$. As the temperature approaches $T_N$ the signal begins to decrease. For higher powers, the onset shifts to lower temperatures due to THz heating of the sample. Dashed white line is a cubic fit to the temperature shift. **b** shows the line-outs for the three indicated powers in **a**. The high power signal is reproduced with an 85 K offset to demonstrate the heating effect. The solid black line shows $R_\perp$ from Figure 1c, indicating how the anisotropic signal changes with temperature.

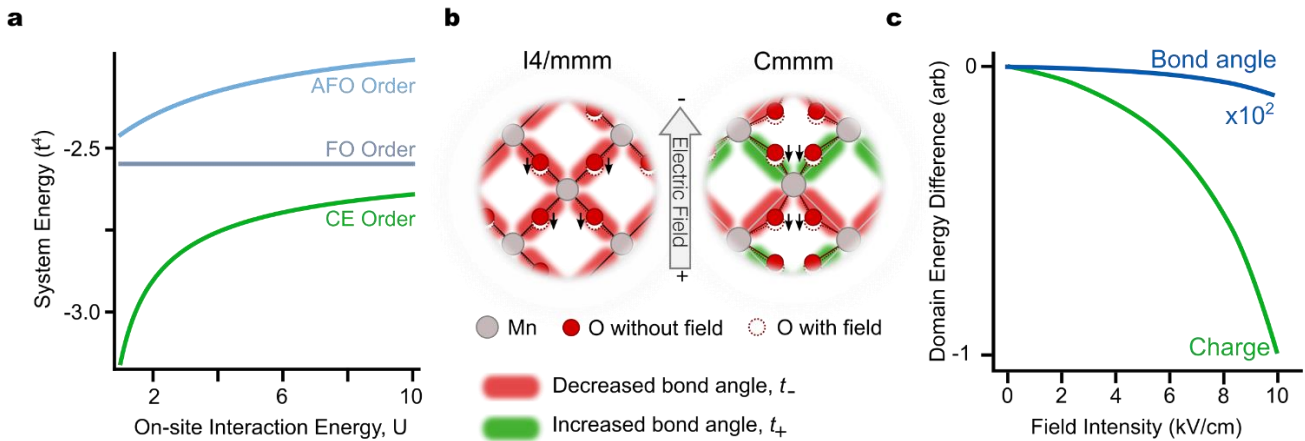

**Figure 4 | Modelling the coupling between electric fields and orbital domains. a** Calculated energy dependence of 3 types of orbital order using Eqn. 1: ferro-orbital (FO), where all occupied orbital are of the same type, antiferro-orbital (AFO) in which the orbital states alternate along both crystallographic directions, and CE order, where the orbital type varies along one axis (anti-ferro) and is constant along the other (ferro). For a wide range of $U$, the experimentally observed CE order is the lowest energy state, demonstrating that the model can reproduce the correct ground state. Note that the CE phase has two degenerate domains in which the ferro and anti-ferro directions are rotated. **b** Atomic displacements induced by the electric field. Positively charged Mn ions are considered stationary and the relative motion due to the charge disproportionation is neglected. Negatively charged O ions are displaced against the field direction. This displacement changes the Mn-O-Mn bond angle. As the hopping term $t$, in Eqn. 1, is proportional to the cosine squared of the Mn-O-Mn bond angle, being largest when $\theta = 180°$, the hopping decreases for smaller bond angles and increases for larger angles. In the I4/mmm symmetry, the Mn-O-Mn bonds are undistorted ($\theta = 180°$) the bond angle is uniformly decreased. In the Cmmm structure the bonds start distorted, which enables some bond angles to increase in the field, improving the hopping. **c** Domain energy splitting when considering the modulation Mn-O-Mn bond angle (blue line) and the field induced effect. Both models reproduce the experimentally observed scaling with $E^2$ but the field induced effect stronger by three orders of magnitude.

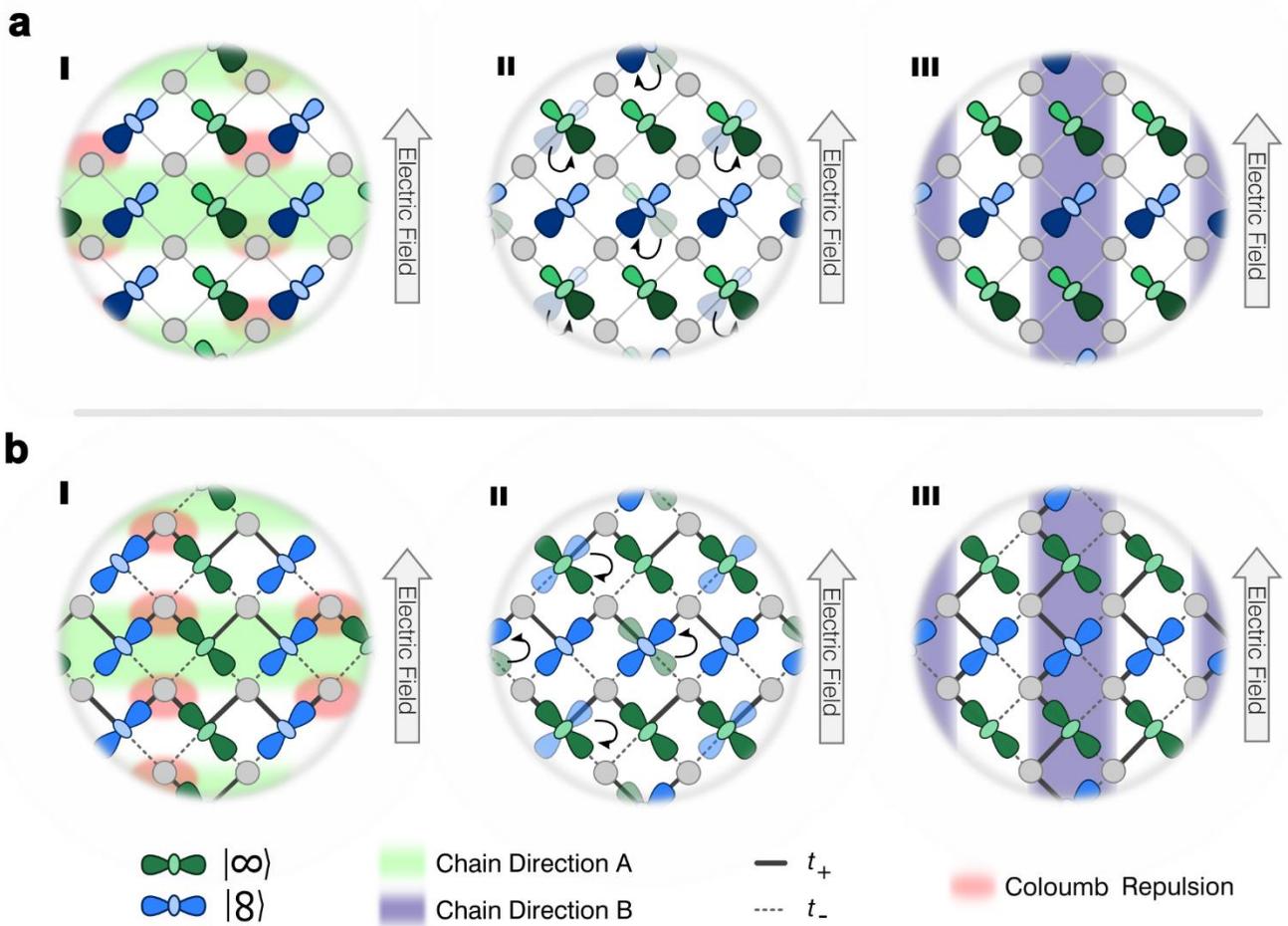

**Figure 5 | Schematics of the two Domain Alignment Scenarios. a** The charge driven process. **I** When the field is applied perpendicular direction of the domain chain, the field causes charge from different $Mn^{3+}$ sites to move towards the same unoccupied $Mn^{4+}$ ion. This has an increased energy cost due to the Coulomb repulsion between the charges. **II**. The energy penalty can cause electrons to change their orbital state so that each electron hops onto a different $Mn^{4+}$ ion. **III**. In the flipped state the orbital domain has been rotated. **b** The structural process. **I** When the field is applied along the chain direction, the Mn-O-Mn bond angle is increased and reduced asymmetrically. As a result hopping is increased such that electrons localized on different $Mn^{3+}$ ions are more likely to hop onto the same $Mn^{4+}$ site, again increasing the energy penalty in the field. **II** Again, the Coulomb energy penalty can be minimized when the orbital rotates and the new orbital pattern (**III**) is a rotated domain. Note that the field polarized along the crystallographic axis does not induce a domain reorientation for either mechanism.

## SUPPLEMENTARY NOTE 1 MODEL AND PERTURBATION EXPANSION

The complex interplay of spin, charge, orbital and lattice degrees of freedom results in fascinating phenomena in manganite compounds such as *e.g.* colossal magneto-resistance (CMR) and various types of orbital ordering[1]. The CE phase of half-doped manganites with perovskite crystal structure, such as LSMO, is a particularly intricate manifestation of these interactions in manganites, that has attracted much theoretical attention due, in part, to its relevance to the CMR effect[2,3]. In the magnetically ordered phase, it consists of neighbouring ferromagnetically ordered zigzag chains stacked antiferromagnetically. This spin pattern is accompanied by two-sublattice charge ordering and occupation of directed orbitals at occupied Mn$^{3+}$ sites along the zigzag chains. The origin of the CE phase has been the subject of much theoretical debate resulting in various proposals to explain various aspects of this peculiar ordering phenomenon based on: double exchange (kinetic energy gain along the ferromagnetic chains) associated with anisotropy of the hopping integrals of e$_g$ orbitals[4–6], the Coulomb interaction[7–11] and electron-lattice coupling[12–18] including both non-cooperative and cooperative Jahn-Teller effects. However, in many of these cases, magnetic order is either assumed or generated via antiferromagnetic super-exchange interactions, in order to obtain the associated orbital ordering. The residual CE-type insulating, orbital order above the Neel temperature may thus be presumably driven by the competition between kinetic energy, and electronic interactions and/or structural distortions[7,19–21] although a different mechanism based on short-range antiferromagnetic correlations has also been proposed[22].

The transition metal oxides are strongly interacting many body systems that, are in practice, modelled by generalized Hubbard models, incorporating competing interactions between the relevant degrees of freedom[23–25]. In order to obtain a qualitative picture of the energetics behind the THz electric-field induced switching of CE orbital domains, we consider a phenomenological model for the paramagnetic phase of 50% hole-doped manganites. We use a simple, purely electronic, extended Hubbard model with two orbitals per site, describing the active e$_g$ orbitals $|\infty\rangle \sim d_{3x^2-r^2}$ and $|8\rangle = d_{3y^2-r^2}$ directed in the **x** and **y** directions respectively, to reproduce the type of observed orbital domains in a single square layer of the material:

$$H = -t\left(\sum_i c^\dagger_{i,\infty} c_{i+\mathbf{e}_x,\infty} + c^\dagger_{i,8} c_{i+\mathbf{e}_y,8} + h.c.\right) + U\sum_i n_{i,\infty} n_{i,8} + V\sum_{<i,j>}(n_{i,\infty} + n_{i,8})(n_{j,\infty} + n_{j,8}) \quad (1)$$

where *U* and *V* are the on-site and nearest neighbour Coulomb interaction terms respectively, with $n_{i,\alpha}$ denoting the electron occupation in the active orbital $\alpha = x, y$ on site *i*. The spatial anisotropy of the *d*-orbitals plays an important role in the properties of the manganites and results in different hopping terms in the **x** and **y** directions (see e.g. ref 3).

In the Hamiltonian above, we make an approximation and assume only anisotropic hopping between the same orbitals such that electrons in the *x* (*y*) orbital hop along the *x* (*y*) -direction. The e$_g$ spins are also coupled to localized core t$_{2g}$ spins which can be modelled via the Hund's coupling[3]. In the strong Hund's coupling limit and assuming the t$_{2g}$ spins to be classical, this interaction induces the alignment of the e$_g$ spin with the core spin which can be effectively described to yield a bandwidth narrowing factor. In the paramagnetic phase, the core spin is averaged out leading to a spinless fermion model with effective hopping (including the band narrowing factor) denoted here by $t$.

The main effect of the THz electric field on the orbitally ordered states can be understood by assuming a stationary field linearly polarized in the plane of the LSMO layer. This is a valid approximation as the THz field frequency is at least an order of magnitude smaller than the hopping dynamics time scales. We describe the electric field $\boldsymbol{E} = E(\boldsymbol{e_x} + \boldsymbol{e_y})$ in the length gauge and consider it, for simplicity, to be polarized at an angle $\pi/4$ with respect to the lattice basis vectors:

$$H_E = -\boldsymbol{E} \cdot \sum_i \sqrt{2} \boldsymbol{r}\, n_i \qquad (2)$$

where $n_i$ is the total electron occupation at site $i$.

The model Hamiltonians Eqs. 1 and 2 are of the form $H = T + H_0$ where $H_0$ consists of terms diagonal in the Fermi occupation number operators $n_{i,\alpha}$. In manganite compounds, the hopping amplitude is much smaller than the interaction energy scales $t \ll U, V$ (ref. 3) and is treated as a perturbation below. The subspace of insulating, charge ordered states with one electron per site localized on a sublattice is the unperturbed ground state subspace at 50% hole doping, in the absence of the electric field **E**.

Since the interaction terms do not distinguish between orbital configurations, this subspace is exponentially degenerate. We consider the effect of the hopping $T$ on this subspace both for $E = 0$ and electric field polarized along the diagonal of the square lattice $E \neq 0$. These can be treated on the same footing as the electric field term, which is diagonal in the occupation number operators, commmutes with the interaction terms.

The effective Hamiltonian up to fourth order in the perturbation $T$ is given by[26,27]:

$$H_2 = P_0 T \frac{1-P_0}{E_0-H_0} T P_0 \qquad (3)$$

$$H_4 = P_0 T \frac{1-P_0}{E_0-H_0} T \frac{1-P_0}{E_0-H_0} T \frac{1-P_0}{E_0-H_0} T P_0$$

$$-\frac{1}{2}\left(P_0 T \frac{1-P_0}{(E_0-H_0)^2} T P_0 T \frac{1-P_0}{E_0-H_0} T P_0 + P_0 T \frac{1-P_0}{E_0-H_0} T P_0 T \frac{1-P_0}{(E_0-H_0)^2} T P_0\right) \qquad (4)$$

where $P_0$ is the projector on to the mentioned ground state subspace. The second order correction $H_2$ describing virtual hopping of an electron on to a neigbouring empty site and back does not distinguish between orbitals occupied. Similarly, the second term in Eq. (4) describes two such sequential processes and also leads to a correction independent of the orbitals involved. Orbital independent corrections do not lift the orbital degeneracy and are henceforth ignored.

Note that since the allowed hopping direction of an electron is determined by the occupied orbital in our model, orbitals cannot exchange places under the action of $T$. Therefore the first term in $H_4$ yields only diagonal energy corrections to orbital configurations and one obtains a classical model of interacting orbitals. This perturbative term describes sequential hopping processes through intermediate excited states. The first hopping process always results in a transfer to an excited state with the charge gap $3V \pm E$ where $\pm$ determines whether the electron hopping is against or along the direction of the field. In the second hopping, an electron may hop, if allowed, onto an already occupied site resulting in an energy contribution $U$, or the second electron in the considered pair hops to an excited state. At this stage the electric field contribution to the excitation energy may be 0 if the two hopping processes occur in opposite directions with respect to the field or $\pm 2E$ if both hops occur against or with the field. The resultant interaction energies of various configurations are produced above (see Supplementary Table 1).

$$A(E) = -t^4 \left( \frac{1}{(3V-E)(U-2E)(3V-E)} \times 1 + \frac{1}{(3V-E)(5V-2E)(3V-E)} \times 4 \right.$$

$$+ \frac{1}{(3V-E)5V(3V-E)} + \frac{1}{(3V-E)5V(3V+E)} + \frac{1}{(3V+E)5V(3V+E)} + \frac{1}{(3V+E)5V(3V-E)}$$

$$+ \frac{1}{(3V+E)(6V+2E)(3V+E)} \times 4$$

$$\left. + \frac{1}{(3V-E)6V(3V-E)} + \frac{1}{(3V-E)6V(3V+E)} + \frac{1}{(3V+E)6V(3V+E)} + \frac{1}{(3V+E)6V(3V-E)} \right)$$

$$B(E) = -t^4 \left( \frac{1}{(3V-E)(5V-2E)(3V-E)} \times 4 + \frac{1}{(3V+E)(5V+2E)(3V+E)} \times 4 \right.$$

$$\left. + \frac{1}{(3V-E)6V(3V-E)} + \frac{1}{(3V-E)6V(3V+E)} + \frac{1}{(3V+E)6V(3V+E)} + \frac{1}{(3V+E)6V(3V-E)} \right) \quad (5)$$

$$C(E) = -t^4 \left( \frac{1}{(3V-E)(6V-2E)(3V-E)} \times 4 + \frac{1}{(3V+E)(6V+2E)(3V+E)} \times 4 \right.$$

$$\left. + \frac{2}{(3V-E)6V(3V-E)} + \frac{2}{(3V-E)6V(3V+E)} + \frac{2}{(3V+E)6V(3V+E)} + \frac{2}{(3V+E)6V(3V-E)} \right) \quad (7)$$

$$Z^{(1)}(E) = -t^4 \left( \frac{1}{(3V-E)(4V+U)(3V-E)} + \frac{1}{(3V-E)(6V+U)(3V+E)} \right.$$

$$+ \frac{1}{(3V+E)(4V+U)(3V-E)} + \frac{1}{(3V-E)(4V+U)(3V+E)}$$

$$\left. + \frac{1}{(3V+E)(5V-2E)(3V-E)} \times 4 + \frac{1}{(3V+E)(5V+2E)(3V-E)} \times 4 \right. \quad (8)$$

$$\left. + \frac{1}{(3V-E)6V(3V-E)} + \frac{1}{(3V-E)6V(3V+E)} + \frac{1}{(3V+E)6V(3V+E)} + \frac{1}{(3V+E)6V(3V-E)} \right)$$

$$Z^{(2)}(E) = -t^4 \left( \frac{1}{(3V-E)(4V+U-2E)(3V-E)} \times 4 + \frac{1}{(3V-E)(6V+U)(3V+E)} \times 4 \right. \quad (9)$$

$$\left. + \frac{2}{(3V-E)5V(3V-E)} + \frac{2}{(3V-E)5V(3V+E)} + \frac{2}{(3V+E)5V(3V-E)} + \frac{2}{(3V+E)5V(3V+E)} \right)$$

$$D(E) = -t^4 \left( \frac{1}{(3V-E)(5V-2E)(3V-E)} \times 4 + \frac{1}{(3V-E)(5V+2E)(3V+E)} \times 4 \right.$$

$$+ \frac{1}{(3V-E)4V(3V-E)} + \frac{1}{(3V-E)4V(3V+E)} + \frac{1}{(3V+E)4V(3V-E)} + \frac{1}{(3V+E)4V(3V+E)}$$

$$\left. + \frac{1}{(3V-E)6V(3V-E)} + \frac{1}{(3V-E)6V(3V+E)} + \frac{1}{(3V+E)6V(3V-E)} + \frac{1}{(3V+E)6V(3V+E)} \right)$$

The checkerboard ordered phase defines a new square lattice rotated by $\pi/4$ radians where, (10) in fourth order perturbation, each site now interacts with its nearest as well as next nearest neighbours with configuration energies given by $D, Z^{(1)}, Z^{(2)}$ and $A, B, C$ respectively. Without the electric field, the orbital degeneracy is lifted as can be seen by comparing energies for orbitally ordered states that are typically in competition in doped manganites: the experimentally observed CE type phase corresponding to the ordering vector $(\frac{\pi}{2}, \frac{\pi}{2})$ in the original lattice, the FO "ferro-orbital" state where all occupied orbitals are of the same type and the "anti-ferro-orbital" state where each line of the original lattice is filled with only one type of orbital and neighbouring lines have orthogonal orbitals. The energies per site for these configurations follow from the energies of the 8 bonds around a given occupied site:

$$\varepsilon_{CE} = (2Z^{(1)}(0) + 2D(0) + 4A(0))/4 \tag{11}$$

$$\varepsilon_{FO} = (4D(0) + 2B(0) + 2C(0))/4 \tag{12}$$

$$\varepsilon_{AFO} = (4Z^{(1)}(0) + 2B(0) + 2C(0))/4 \tag{13}$$

For a wide window of values $\frac{U}{V} > 1$, the CE type state is seen to be the lowest energy state (see Fig.4 in the main text, with set energy scale $V > 1$). In fact, the closest competing state is the FO state which becomes favorable above the critical value $U \approx 87.3734V$. We have also checked, using exact diagonalization of an 8-site cluster, that the CE type phase is indeed the ground state of the model Eq. (1) in the limit $t \ll U, V$ with $U \sim V$.

## Supplementary Table 1 Effective Orbital Interactions

Effective interactions between orbitals corresponding to different charge configurations on the lattice, as obtained in fourth order perturbation expansion. The corresponding expressions are presented in Eqs.(5-10). The next nearest neighbours in the **x, y** directions are collected in the first row, while the diagonal neighbours appear in the lower row.

| $(\infty \ 0 \ 8), \begin{pmatrix}\infty\\0\\8\end{pmatrix} : A(\mathbf{E})$ | $(\infty \ 0 \ 8), \begin{pmatrix}\infty\\0\\8\end{pmatrix} : A(-\mathbf{E})$ | $(\infty \ 0 \ \infty), \begin{pmatrix}8\\0\\8\end{pmatrix} : B(\mathbf{E})$ | $(8 \ 0 \ 8), \begin{pmatrix}\infty\\0\\\infty\end{pmatrix} : C(\mathbf{E})$ |
|---|---|---|---|
| $\begin{pmatrix}0 & 8\\ \infty & 0\end{pmatrix}, \begin{pmatrix}0 & \infty\\ 8 & 0\end{pmatrix} : Z^{(1)}(\mathbf{E})$ | $\begin{pmatrix}\infty & 0\\ 0 & 8\end{pmatrix} : Z^{(2)}(\mathbf{E})$ | $\begin{pmatrix}8 & 0\\ 0 & \infty\end{pmatrix} : Z^{(2)}(-\mathbf{E})$ | $\begin{pmatrix}0 & \infty\\ \infty & 0\end{pmatrix} : D(\mathbf{E})$ |

## Supplementary Note 2 Effect of electric field on CE type domains

Having obtained a description of the CE type phase, we turn now to the effect of the electric field on the degenerate domain A and domain B states. The validity of the perturbation expansion above is based on assuming that the electric field is off resonant with the interaction energies so that the ideal checker-board charge ordered space remains energetically well separated from other possible charge ordered states. The energies per site in domains A and B are then

$$\varepsilon_{CE_A}(E) = \frac{2Z^{(1)}(E) + 2D(E) + 2A(E) - 2A(-E)}{4} \tag{14}$$

$$\varepsilon_{CE_B}(E) = (Z^{(2)}(E) + Z^{(2)}(-E) + 2D(E) + 2A(E) - 2A(-E))/4 \tag{15}$$

The inequivalence of the zigzag chains, i.e. the nearest neighbour interactions in the rotated lattice along the chains, determine the energy level splitting

$$\Delta\varepsilon = \varepsilon_{CE_A}(E) - \varepsilon_{CE_B}(E) = (Z^{(2)}(E) + Z^{(2)}(-E) + 2D(E) + 2A(E) - 2A(-E))/4 \tag{16}$$

For weak fields $E \ll U, V$, this splitting grows quadratically with the electric field, i.e. linearly with the intensity. This is a manifestation of the invariance of the ground state configuration with respect to reversal of the electric field.

$$\Delta\varepsilon \to -2t^4 \frac{(476U^3 + 5337U^2V + 15348UV^2 - 7036V^3)}{30375V^5(U+4V)^3} E^2 \qquad (17)$$

The domain A configuration, with zigzag chains aligned along the direction of the field thus becomes energetically more favourable than domain B and represents a more stable thermodynamic phase. Therefore, in the experimental situation, wrongly aligned domains should be expected to undergo rotation of orbitals to the energetically favourable alignment.

In the above, we have not taken into account structural distortions and the corresponding electron-lattice interactions, since the THz field couples weakly and non-resonantly to the phonon degrees of freedom (however see section below). Thus, for simplicity, we have also aimed for a description of the field-less phases purely in terms of electronic degrees of freedom using a simplified hopping matrix Eq. (1) which yields the CE-type phases. A more realistic model would include the effects of non-orthogonality of the elongated orbitals $d_{3x(y)^2-r^2}$, and can be most succinctly captured by the hopping matrix in terms of an orthogonal $e_g$ basis given by Slater-Koster integrals (see *e.g.* ref. 3 for a comprehensive review of theoretical approaches to magnetically ordered manganites incorporating the interplay of spin, orbital and charge degrees of freedom). In this case, in the paramagnetic phase, the CE orbital ordered phase with elongated orbitals at the Mn$^{3+}$ sites has, in fact, been shown to be stabilized by anharmonic contributions to the elastic energy associated with structural distortions[7,21]. Thus our approximate approach captures the essential qualitative features of a more realistic model for CE type orbital ordering. Similarly, the effect of the electric field E on the relative stability of the states corresponding to the two kinds of domains will be qualitatively similar in the two models. Indeed, even though the perturbative corrections would include contributions due to the gain in elastic energy, as well as new hopping amplitudes, the electric field energetically favours hopping along the field which will result in a splitting in energies of the two domain states. Furthermore, since the physics is symmetric with respect to the inversion $E \to -E$, the lowest order splitting is $\sim E^2$.

## SUPPLEMENTARY NOTE 3 ELECTRIC FIELD INDUCED STRUCTURAL CHANGES

As an alternative to the above purely electronic scenario, we consider the effect of electric field induced static lattice distortion on the two types of CE domain configurations. This lattice distortion can be taken into account primarily as a change of the hopping amplitudes $t$. Already in the absence of the electric field, the so-called tolerance factor in manganite compounds quantifies a structural distortion which can lead to the deviation of the Mn-O-Mn bond angle from the value $\pi$ corresponding to ideal cubic symmetry[1,3]. An electric field acting at angle $\pi/4$ with respect to the square lattice basis vectors leads to the dimerization of the hopping Hamiltonian $T$. Nearest neighbours on the square lattice connected by the same effective amplitude form zigzag chains. Two neighbouring zigzag chains are characterized by different hopping amplitudes $t_1$ and $t_2$ (see Fig. 4 in main text). The Hamiltonian thus obtained has the same structure as in Eq. (1) but now with the described dimerized hopping. As in the previous section, fourth order perturbation theory can be invoked to obtain effective interactions between orbitals in the charge ordered subspace. The hopping dimerzation leads to an energy splitting (per site) between the two types of CE type zigzag orbital domains A and B

$$\Delta\varepsilon_{\rm d} = 2t^4 \frac{(t_1^2 - t_2^2)^2 (7U - 2V)}{270V^3(U+4V)} \qquad (18)$$

To model the field dependence on the structure, we assume that only the oxygen ions move relative to the manganese ions which makes a small change in the Mn-O-Mn bond angle. The effective hopping amplitude $t$ can be expressed as $t = t_0 \cos^2(\pi - \theta_0)$, where $t_0$ is the hopping integral for a flat Mn-O-Mn bond[28]. Small changes in the bond angle, $\theta_0$, change the hopping as $t' = t(1 + 2\Delta\theta \tan\theta)$ and we neglect effects arising from the movement of the oxygen ion along the Mn-Mn bond direction. When bond angles that determine $t_1$ and $t_2$ change in the opposite direction we have $(t_1^2 - t_2^2)^2 = 64(\Delta\theta \tan\theta)^2$.

In order to compare Eq. (18) with Eq. (17) we need to express $\Delta\theta$ as a function of the electric field. For a small displacement $\Delta x$ of the oxygen ion away from the Mn-Mn bond axis results in a change in bond angle give as $\Delta\theta = 4\Delta x/r_{MM}$, where $r_{MM}$ is the Mn-Mn distance. The relation between $\Delta x$ and the electric field can then be obtained by considering the response of a dipole displacement driven by a non-resonant DC field as $qE = -m\omega_0^2 \Delta x$, where $q, m, \omega_0$ are the dipole charge, mass and resonant frequency respectively. This then gives the following,

$$\Delta\theta = \frac{4qE}{\sqrt{2}m\omega_0^2 r_{MM}} = \alpha E, \tag{19}$$

where the factor of $\sqrt{2}$ comes from the fact that the electric field is at an angle $\pi/4$ to the displacememnt. Equation (18) can then be used to calculate the change in $t$. In the paper we assume that the dipole charge is approximately the electron charge $q \approx e = 1.6 \times 10^{-19}$C, the dipole mass can be approximated by the oxygen mass $m = m_O = 2.7 \times 10^{-26}$ kg, the restoring force is given by the lowest frequency IR active phonon mode at $\omega_0 = 2\pi \times 10^{-12}$ s$^{-1}$. Although these quantities are estimates they should be correct to within an order of magnitude. Finally, $r_{MM} = 3.9 \times 10^{-10}$m and $\theta_0 = 176.72°$ are obtained from diffraction data of LSMO in the $C_{mmm}$ phase[29].

## Supplementary References